\documentclass{pasj01}
\draft 
\Received{$\langle$reception date$\rangle$}
\Accepted{$\langle$acception date$\rangle$}
\Published{$\langle$publication date$\rangle$}
\usepackage{color}
\usepackage{ulem}

\begin{document}

\title{High-mass star formation in Orion possibly triggered by cloud-cloud collision III; NGC~2068 and NGC~2071}
\author{Shinji FUJITA$^{\rm 1, *}$, Daichi TSUTSUMI$^{1}$, Akio OHAMA$^{1}, $\textcolor{black}{Asao HABE$^{2}$, Nirmit SAKRE$^{2}$}, Kazuki OKAWA$^{1}$, Mikito KOHNO$^{1}$, Yusuke HATTORI$^{1}$, Atsushi NISHIMURA$^{3}$, Kazufumi TORII$^{4}$, Hidetoshi SANO$^{1}$, Kengo TACHIHARA$^{1}$, Kimihiro KIMURA$^{1}$, Hideo OGAWA$^{3}$, and Yasuo FUKUI$^{1}$}

\altaffiltext{1}{Department of Physics, Nagoya University, Furo-cho, Chikusa-ku, Nagoya, Aichi 464-8602, Japan}
\altaffiltext{2}{Department of Physics, Graduate school of Science, Hokkaido University, Kita 10 Nishi 8 Kita-ku, Sapporo 060-0810, Hokkaido, Japan}
\altaffiltext{3}{Department of Physical Science, Graduate School of Science, Osaka Prefecture University 1-1 Gakuen-cho, Naka-ku, Sakai, Osaka 599-8531, Japan}
\altaffiltext{4}{Nobeyama Radio Observatory, National Astronomical Observatory of Japan (NAOJ), National Institutes of Natural Sciences (NINS), 462-2 Nobeyama, Minamimaki, Minamisaku, Nagano 384-1305, Japan}

\email{fujita.shinji@a.phys.nagoya-u.ac.jp}


\maketitle

\begin{abstract}

Using the NANTEN2 Observatory, we carried out a molecular line study of high-mass star forming regions with reflection nebulae, NGC~2068 and NGC~2071, in Orion in the $^{13}$CO($J$=2--1) transition.
The $^{13}$CO distribution shows that there are two velocity components at \textcolor{black}{9.0} and \textcolor{black}{10.5} km s$^{-1}$. 
The blue-shifted component is in the northeast associated with NGC~2071, \textcolor{black}{whereas} the red-shifted component is in the southwest associated with NGC~2068.
\textcolor{black}{The total intensity distribution of the two clouds shows a gap of $\sim$ 1 pc,} suggesting that they are detached at present.
A detailed spatial comparison 
indicates that the two show complementary distributions.
The blue-shifted component lies toward an intensity depression to the northwest of the red-shifted component, where we find that a displacement of \textcolor{black}{0.8} pc makes the two clouds fit well with each other.
\textcolor{black}{Furthermore, a new simulation of non-frontal collisions shows that observations from 60$^\circ$}
\textcolor{black}{off the collisional axis}
\textcolor{black}{agreed well with the velocity structure in this region.}
\textcolor{black}{On the basis of} these results, we hypothesize that the two components collided with each other at a projected relative velocity \textcolor{black}{3.0} km s$^{-1}$.
The timescale of the collision is estimated to be \textcolor{black}{0.3} Myr for an assumed \textcolor{black}{axis} of the relative motion 60$^\circ$ off the line of sight.
We assume that the two most massive early B-type stars in the cloud, illuminating stars of the two reflection nebulae, 
were formed by 
collisional triggering at the interfaces between the two clouds.
\textcolor{black}{Given} the other young high-mass star forming regions, \textcolor{black}{namely}, M42, M43, and NGC~2024 (\textcolor{black}{Fukui et al. 2018b}; Ohama et al. 2017a), it seems possible that collisional triggering \textcolor{black}{has been} independently working to form O-type and early B-type stars in Orion in the last Myr over a projected distance of $\sim$ 80 pc. 
\end{abstract}

\section{Introduction}
\subsection{Background}
High-mass stars are \textcolor{black}{highly} influential \textcolor{black}{in exciting} motions of the interstellar medium (ISM) and \textcolor{black}{enriching} metals in the ISM.
It is therefore of crucial importance to better understand the mechanism of high-mass star formation in order to elucidate Galaxy evolution. 
The Orion region includes various types of star formation and has been \textcolor{black}{an ideal} test bench for theories on high-mass star formation mechanisms at \textcolor{black}{a close} distance \textcolor{black}{from} the sun around 400 pc.
The region hosts young star formation including H{\sc ii} regions, reflection nebulae and massive star clusters, e.g., M42, M43, NGC~2023, NGC~2024, NGC~2068, and NGC~2071.
\textcolor{black}{The region also hosts} more evolved stars in OB associations without molecular gas (see Bally (2008) for a review and references therein).
The stellar ages range from less than 1 Myr to 10 Myr.
The OB associations \textcolor{black}{are understood as showing} an age sequence in space, in the sense that Ori OB1a in the north is the oldest, whereas Ori OB1d, \textcolor{black}{ lying in the direction of M42, is the youngest} (Blaauw 1991).
It \textcolor{black}{has therefore been} speculated that some common mechanism may be working to form OB associations at a semi regular separation of 25 pc.
In the 1970s, a star formation scenario was proposed to explain the distribution of OB associations.
Elmegreen \& Lada (1977) presented a model of sequential formation of OB associations based on a shock compressed layer driven by an ionization-shock front of H{\sc ii} regions.
Their model explained formation of the quasi-regularly spaced OB associations with an age sequence in a large giant molecular cloud extended along the galactic plane \textcolor{black}{as the mass reservoir}. 
\textcolor{black}{This scenario was later developed as a ``collect-and-collapse'' scenario in smaller H{\sc ii} regions including the Spitzer bubbles, and a number of theoretical and observational studies were made on second generation star formation triggered by O-type stars in the bubbles (e.g., Deharveng et al. 2006; Dale et al. 2007; Walch et al. 2015).}
However, it seems that these scenarios require additional elaboration by confrontation between theories and observations.
 \textcolor{black}{For example}, it remains unexplained  \textcolor{black}{whether} the first generation O-type star  \textcolor{black}{or stars} were formed by a spontaneous process. 

\subsection{Observed cloud-cloud collisions}
\textcolor{black}{Molecular cloud observations have been} analyzed and interpreted under the assumption that an apparently single-peaked cloud is a single component governed by cloud's self-gravity \textcolor{black}{(Zinnecker \& Yorke 2007, Tan et al. 2014, Egan et al. 1998)}.
This is a natural \textcolor{black}{assumption, since} a molecular cloud often appears self-gravitating with a single velocity component.
Averaged kinetic properties like velocity dispersion have been studied in depth in numerous papers\textcolor{black}{(e.g., McKee \& Ostriker 2007 and references therein)} to test the virial theorem using Larson's law (Larson 1981).
\textcolor{black}{Many observational studies of turbulence, infall motion, and outflow in the molecular cloud are often conducted by using molecular emission lines, whereas studies of the detailed spatial structure and the physical state of the cloud are often conducted using dust continuum emissions.}
\textcolor{black}{Dust emission data  have high angular resolutions but do not include velocity information about the cloud.}
Meanwhile, in super star clusters \textcolor{black}{(SSCs)}, pairs of velocity clouds whose velocity separation is supersonic (10--15 km s$^{-1}$ or more) \textcolor{black}{have been} found in Westerlund2 \textcolor{black}{($\sim$ 16 km s$^{-1}$, Furukawa et al. 2009)}, NGC~3603 \textcolor{black}{($\sim$ 20 km s$^{-1}$, Fukui et al. 2014)}, RCW38 \textcolor{black}{(14 km s$^{-1}$, Fukui et al. 2016)}, and R136 \textcolor{black}{in \textcolor{black}{the} Large Magellanic Cloud ($\sim$ 70 km s$^{-1}$, Fukui et al. 2017).}
These clusters are all associated with nebulosity and intense dust emission, unlike the rest of the \textcolor{black}{SSCs} in the Milky Way, suggesting that these four are the youngest among the \textcolor{black}{SSCs} (Portegies Zwart et al. 2010; Fukui et al. 2016, 2017a; Kuwahara et al. 2018 in preparation).
\textcolor{black}{The suggested interpretation for these SSCs is} that cloud-cloud collision (CCC) took place, triggering formation of a cluster in the collisional shock-compressed layer (Inoue \& Fukui 2013).
The large velocity separation cannot be gravitationally bound by cloud gravity, so chance collision between two giant molecular clouds 
is a plausible scenario.
\textcolor{black}{The high nebulosity of these four SSCs suggests that CCC may be an important process in massive SSC formation.}
\textcolor{black}{However it may be difficult to detect molecular clouds associating with SSCs long after their formation, as the parent cloud may have since been dispersed by the stellar feedback.}
If two clouds are observed clearly separated in velocity and are connected by bridge features, collision is well corroborated. 
Such a rare example is found in RCW38\textcolor{black}{, which shows} two components with 12 km s$^{-1}$ velocity separation, connected by a bridge feature within 1 pc of the cluster center (Fukui et al. 2016).
A clear case like RCW38 is, however, not often seen, because its age of $10^5$ yr is considerably younger than the typical age of clusters, i.e., more than 2 Myr.

Subsequently, formation of a single O-type star was found to be triggered by CCC in other systems; e.g., M20, RCW120, N159W and N159E (Torii et al. 2011; Torii et al. 2015; Fukui et al. 2015; Saigo et al. 2017).
These cases show a variety of velocity separation from 2 to 20 km s$^{-1}$, and in the latter two cases the velocity separation in projection is only a few km s$^{-1}$. 
\textcolor{black}{These authors have} shown that the two colliding clouds physically merge with each other by collision and are often observed as a single cloud, \textcolor{black}{particularly} if one of the clouds is dominant in column density.
This possibility, \textcolor{black}{of a colliding cloud pair that is} apparently single peaked, is illustrated by synthetic observations of colliding clouds by \textcolor{black}{Fukui et al. (2018b)} based on numerical simulations by Takahira et al. (2014).
Such cloud pairs may not be separable by observed velocity, even when supersonic CCC is taking place.
Since supersonic collision can significantly influence cloud physical states and star formation via shock compression (Inoue \& Fukui 2013), a careful investigation of cloud kinematics is required in order to capture collision signatures and to gain insight into the collision physics.

Theoretical simulations at the Galaxy scale show that collisions between molecular clouds are frequent in the Galactic disk (Fujimoto et al. 2014; Dobbs et al. 2015).
An important question is whether CCC is a common process in O-type star formation or an adhoc scenario that rarely happens.
In light of the extreme importance of O-type star formation in astrophysics, it is an urgent issue to quantitatively explore the role of CCC in high-mass star formation.

\subsection{Theories of collision}
CCC offers an alternative to ionization-driven shock compression in O-type star formation as shown by numerical simulations.
Numerical hydrodynamic simulations by Takahira et al. (2014) showed a simple case of collision in which a small cloud collided with a large cloud to produce a cavity in a large cloud (see also Habe and Ohta 1992; Anathpindika 2010).
Note that collision between similar-sized clouds \textcolor{black}{was found to be} rarer than \textcolor{black}{that} between dissimilar\textcolor{black}{-sized} clouds.
These results were used to synthesize observations on a cloud scale.
\textcolor{black}{These authors then showed} that a complementary distribution between the two clouds was a common observable signature, as long as ionization by the formed star is not destroying the clouds \textcolor{black}{(Fukui et al. 2018b)}.
It was also shown that \textcolor{black}{a pair of} colliding clouds merge into a single peak and is not usually seen as two distinct components due to gas in the intermediate velocity range between the two components.

On a microscopic scale, the magnetohydrodynamic simulations by Inoue \& Fukui (2013) showed that the interface layer between the two clouds became strongly compressed.
The particles in the interface layer showed large velocity differences, and the collisional interaction caused a large velocity dispersion, of magnitude similar to the actual velocity difference between the two clouds.
The layer must be highly \textcolor{black}{heterogeneous} with clumps, because the initial density distribution prior to collision is highly \textcolor{black}{heterogeneous;} this is a general ISM property.
The clumps amplify turbulence and magnetic field by deflection of shock fronts, and realize the high-mass accretion rate \textcolor{black}{(10$^{-3}$ to 10$^{-4}$ M$_\odot$ yr$^{-1}$)} required to form high-mass stars (Inoue \& Fukui 2013).

\subsection{NGC~2068 and NGC~2071}
Large-scale molecular observations covering the Orion star-forming region have been made by a number of authors mainly in CO emission. 
\textcolor{black}{These observations} have been used to address the star formation process (e.g., Sakamoto et al. 1994; Wilson et al. 2005; Nishimura et al. 2015).
A number of works focusing on infrared sources in NGC~2068 and NGC~2071 in the north of L1630 have been made at various wavelengths and have shown active formation sites of early B-type star and clusters (for the early works, see, e.g., Lada \& Lada 2003 and references is therein).
\textcolor{black}{The Most} recent near infrared observations have selected 186 candidate young stellar objects in the region that show good spatial correlation with optical extinction (Spezzi et al. 2015).
\textcolor{black}{Molecular emission observation was not as intensive for this region as for the M42 region, and a limited number of molecular observations covered molecular gas on a degree scale}
(e.g., Aoyama et al. 2001; Ikeda et al. 2009; Buckle et al. 2010).
For a review of the NGC~2068 and NGC~2071 region, please see Gibb (2008) and references therein.

\subsection{Aims of the present paper}
In the present paper we analyze new $^{13}$CO($J$=2--1) data with  particular attention to the velocity distributions in NGC~2068 and NGC~2071.
Our aim is to test if CCC is a viable scenario in this region by using the analysis applied to M42, M43, NGC~2024, and other regions of high-mass star formation.
We emphasize on utilization of recently developed methods on CCC based on hydrodynamic numerical simulations, as well as empirical signatures acquired in studies of CCC (e.g., \textcolor{black}{Fukui et al. 2018b}).
The paper is organized as follows.
Section 2 describes details of observations with the NANTEN2-Observatory.
Section \textcolor{black}{3} provides the results of the present observations and analysis.
Section \textcolor{black}{4} offers discussion on the results and their implications for high-mass star formation.
Section \textcolor{black}{5} concludes the paper.

\section{Observations}
\textcolor{black}{We carried out observations in $^{12}$CO($J$=2--1) and $^{13}$CO($J$=2--1) emissions over an area of 1 $\times$ 1 deg$^2$ in the direction of} NGC 2068 and NGC 2071 using the NANTEN2 4-m millimeter/sub-millimeter telescope in Atacama, Chile (4850 m) in December 2016. 
The beam size and the velocity resolution were 90$^{\prime \prime}$ and 0.08 km s$^{-1}$ at 230 GHz, respectively.
These observations were made using the on-the-fly mapping technique (Sawada et al. 2008).
The backend was a digital Fourier transform spectrometer with 16384 channels of 1 GHz bandwidth.
After baseline subtraction, we created the FITS image with a 30$^{\prime \prime}$ spatial grid and a 0.08 km s$^{-1}$ velocity grid.
\textcolor{black}{The typical system noise temperature, including atmospheric noise, was 270--340 K for $^{13}$CO($J$=2--1)}, and the typical rms noise levels of the spectral data were 0.6 and 0.7 K, 
\textcolor{black}{scaled in Tmb}, respectively, for $^{12}$CO($J$=2--1) and $^{13}$CO($J$=2--1).
The pointing accuracy was better than 10$^{\prime \prime}$. We used the CO($J$=2--1) data from Orion B obtained by the Osaka Prefecture University 1.85-m telescope (Nishimura et al. 2015) as a standard source \textcolor{black}{for intensity calibration.}
We used the Galactic coordinate system in the present paper.

\section{Results}
Because $^{12}$CO emission is heavily saturated with self-absorption, we mainly used $^{13}$CO data in the present analysis.
\textcolor{black}{Figure 1a} shows the integrated intensity distribution of the $^{13}$CO emission.
The molecular distributions are divided into three components, namely, GC~2068, NGC~2071 and NGC~2071-north, the former two being dominant.
The exciting stars of NGC2068, namely, B-type stars HD~38563 and HD~290862, are located close to a $^{13}$CO peak at ($l, \, b$)=(205\fdg38, -14\fdg32) and the exciting star of NGC~2071, namely, B-type star HD~290861 (V*~V1380~Ori), near a $^{13}$CO peak at ($l, \, b$)=(205\fdg11,-14\fdg11) (Strom et al. 1975).
\textcolor{black}{The $^{12}$CO($J$=2--1) data revealed that two infrared sources (LBS17 and LBS8, Gibb \& Heaton 1993) showed protostellar outflows.}
Iwata et al. (1988) discovered CO outflow at ($l, \, b$)=(204\fdg868, -13\fdg866) in molecular clump NGC~2071-north which seems to be connected with the NGC~2071 cloud in the north of the study region (Figure 1a).
186 Young Stellar Object candidates (YSOc) (Spezzi et al. 2015) were found near ridges of elevated $^{13}$CO($J$=2--1) intensity.
In the blue-shifted cloud, the main molecular ridge of 0.4$^\circ
$ length elongated in the east-west direction in the equatorial coordinate was associated with $\sim$ 50 YSOc, whereas in the red-shifted cloud, the main ridge of 0.4$^\circ
$ length elongated in the north-south direction was associated with another $\sim$ 50 YSOc (\textcolor{black}{Figure 1a}).
\textcolor{black}{Figure 1b shows a composite of the WISE 12 $\mu$m (red), 4.6 $\mu$m (green), and 3.8 $\mu$m (blue) images.
The contour shows the $^{13}$CO distribution as in Figure 1a. 
Because the reflection nebulas are also clearly divided at the same position with regard to $^{13}$CO, the northern and southern clouds appear to be associated with each other.}

\textcolor{black}{Figure 2a is a 45$^\circ$ rotated map of Figure 1a that separates the northern and southern clouds, the solid lines being the integral range of Figure 2b.}
\textcolor{black}{We find that the cloud can be clearly separated also in the position velocities.}
\textcolor{black}{As shown by the black lines in Figure 2b indicating the intensity-weighted mean velocity along the y-axis at every 30$^{\prime\prime}$. 
These two molecular clouds seem to be separated clouds rather than one cloud with a velocity gradient.}
\textcolor{black}{The intensity-weighted mean velocities of the $^{13}$CO emission within the Boxes A and B are 9.0 and 10.5 km s$^{-1}$, respectively. }
We hereafter refer to the northern cloud as the blue-shifted cloud and the southern cloud as the red-shifted cloud.

In order to see greater details, 
\textcolor{black}{we plotted a} velocity-channel distribution of $^{13}$CO($J$=2--1) every \textcolor{black}{0.5 km s$^{-1}$ (Figure 3).}
\textcolor{black}{By considering Figures 2b and 3, we provisionally determined velocity ranges of the blue- and red-shifted clouds as 7.0--9.0 and 10.5--12.5 km s$^{-1}$, respectively, in order to reveal the distributions of the two velocity components.
The intermediate velocity emissions (9.0--10.5 km s$^{-1}$) have been excluded from the velocity ranges because the two components are blended.}

\textcolor{black}{Figures 4a, 4b, and 4c show the spatial distribution of the blue-shifted cloud, the intermediate velocity cloud, and the red-shifted cloud.
The blue-shifted cloud is mainly distributed on the north side and extends to the east-west direction.
Peak location is ($l, \, b$)$\sim$(205\fdg00, -14\fdg2), with a characteristic small peak found at ($l, \, b$)=(205\fdg25, -14\fdg3).
In contrast, the red-shifted cloud is mainly distributed on the south side and is distributed from north to south.
The red-shifted cloud peak is at ($l, \, b$)=(205\fdg4, -14\fdg32) near the B-type star, the driving source of outflow.
The intermediate velocity cloud is distributed throughout, with north- and south-side shapes similar to those of the blue- and red-shifted clouds, respectively.}

\textcolor{black}{By using the assumptions that $^{13}$CO($J$=2--1) is optically thin and in the local thermodynamic equilibrium,} the cloud mass and the column density were calculated to be 1 $\times$ $10^3$ M$_\odot$ and 2 $\times$ $10^{22}$ cm$^{-2}$, respectively, for the blue-shifted cloud at an integrated intensity of 13.5 K km s$^{-1}$.
The cloud mass and peak column density of the red-shifted cloud were estimated to be 2 $\times$ 10$^3$ M$_\odot$ and 6 $\times$ $10^{22}$ cm$^{-2}$, respectively, at an integrated intensity of 10 K km s$^{-1}$.
The excitation temperature T$\rm _{ex}$ was 15 K for the blue-shifted cloud and 30 K for the red-shifted cloud, as derived from the peak brightness temperature of the $^{12}$CO emission.
We assumed [$^{13}$CO]/[H$_2$=2 $\times$ 10$^{-6}$] for estimating of the molecular column density (e.g., Dickman 1978; Frerking et al. 1982).

By analyzing $^{13}$CO($J$=2--1) data, we \textcolor{black}{have} investigated detailed kinematic properties of the star forming molecular gas whose column density lines between 10$^{22}$ and 10$^{23}$ cm$^{-2}$.
The blue-shifted cloud and the red-shifted cloud, are spatially separated by a typical gap of 1 pc in the sky.
The northern cloud has an average velocity of \textcolor{black}{9.0} km s$^{-1}$ and is associated with reflection nebula NGC~2071 and the southern cloud has an average velocity of \textcolor{black}{10.5} km s$^{-1}$ and is associated with reflection nebula NGC~2068.
Each cloud is associated with one or two early B-type stars, a protostellar outflow driven by young high-mass protostars, and several tens of YSOc.

\section{Discussion}

\subsection{A cloud-cloud collision scenario}
\textcolor{black}{As mentioned in Section 1, high-mass star formations triggered by CCC have been reported by many studies toward other H{\sc ii} regions. 
At least two clouds, separated in velocity, are associated with them.
Also in the present study, the two velocity cloud, the blue- and red-shifted clouds, are associated with the two B-type stars HD~38563 and HD~290862.
We built a hypothesis that the two clouds collided and the B-type stars formed in this region. }

\textcolor{black}{The detailed distributions around ($l, \, b$)=(205\fdg3, -14\fdg3) of the blue- and red-shifted clouds are shown in Figure \ref{disp2}a.}
\textcolor{black}{The red-shifted cloud showed two peaks, one at ($l, \, b$)=(205\fdg40, -14\fdg32) and the other at the peak near the two B-type stars.
The two B-type stars seem to be associated with the red-shifted cloud at least. 
In the hypothesis, the two clouds were collided and formed the B-type stars at the dense part in the cloud indicated by the green square in Figure \ref{disp2}a, and the initial position of the clouds are expected as shown in Figure \ref{disp2}b with a shift of 0.8 pc. }

\textcolor{black}{To consider the collision hypothesis, we make a collision simulation as discribed in Takahira et al. (2014).}
\textcolor{black}{Physical parameters of the simulation are listed in Table 1,}
\textcolor{black}{with the two identical clouds placed separated by 5 pc as an initial condition.}
\textcolor{black}{Both clouds are assumed to be spherically symmetric, with radius and density of 2 pc and n$_{\rm H_2}$ = 1.2 $\times$ 10$^3$ cm$^{-3}$, respectively.
These two clouds are assigned an internal turbulence of velocity dispersion of 1.53 km s$^{-1}$ leading to highly heterogenous density distributions.}
\textcolor{black}{In initial conditions, cloud B is at ($X,\: Y,\: Z$)=($+2.5$, 0, 0) pc, and cloud A is at ($X,\: Y,\: Z$)=(-2.5, 0, 0) pc, with a collision velocity ($v_X,\: v_Y,\: v_Z$)=(2.5$\sqrt{3}$, 2.5, 0).
Figure \ref{sim}a shows a three-dimensional schematic diagram of the initial clouds of this simulation.
The spatial distribution of the initial clouds in the X--Y plane is shown in  Figure \ref{sim}b. 
Although Takahira et al. (2014) simulated a head-on collision of large and small clouds, the present simulations deal with oblique collision between two similar clouds.  
Figures \ref{sim}c and \ref{sim}d show three-dimensional schematic diagrams after 0.5 Myr and the gas distribution on the X--Y plane, respectively. 
The projected gas distribution and velocity structure change with the line of sight.}
\textcolor{black}{After 0.5 Myr, the projected gas distribution of the two clouds along a vector of ($X,\: Y,\: Z$)=(-$\sqrt{3}/2, 0, 1/2$) is shown in Figure \ref{sim}e.}
\textcolor{black}{Figure \ref{sim}f shows a position-velocity diagram along the A-axis defined by the solid line in Figure \ref{sim}e.}
\textcolor{black}{The spatial distribution of the two collided clouds looks like a single cloud with two peaks in Figure \ref{sim}e.}
\textcolor{black}{The velocity components of the two clouds mix at the collision interface, whereas, the non-interacting part retains the initial velocity.
Since Figure \ref{sim}e is similar to the actual observation in Figure 2b, our observation is well reproduced by this simulation.
}

\textcolor{black}{Figure \ref{disp1}a shows the entire distributions of the two velocity components.
The shapes of the blue- and red-shifted clouds can be lined up by moving the shape of each boundary surface approximately southward.
In Figure \ref{disp1}b, the displacement of 0.8 pc shown by a green arrow, which is the same displacement in Figure \ref{disp2}b, produces a good complementary fit on the large scale, supporting the notion of physical association between the two components.
For the quantitative evaluating of the cloud displacement, we use Spearman's rank correlation coefficient between the CO intensities of the two clouds for pixel to pixel.
We calculated the correlation coefficient for each displacement of the blue-shifted cloud with the red-shifted cloud; the smaller correlation coefficient (anti-correlation) corresponds to a complementary distribution each other.
Figure \ref{disp1}c shows a map of the correlation coefficients as a function of a displacement of the blue-shifted cloud.
The $x$--axes and $y$--axes correspond respectively to Galactic Longitude displacement and Galactic Latitude displacement, and the white circles indicate the original position of the blue-shifted cloud (without displacement).
The low-value region, corresponding to highly complementary, extends over the range of $\sim$ 1 pc from the east to the west, and the white square indicates the displacement with the smallest value, corresponding to the most complementary.
Because the spatial resolution of the CO data is $\sim$0.2 pc, the estimation of the best displacement by this method may included errors of a few sub-pc at least.
Therefore, there is no significant difference between the two displacement indicated by the white square and the triangle marker in Figure \ref{disp1}c. 
For these reasons, we determined the initial position of the two colliding clouds in this CCC scenario as shown in Figure \ref{disp1}b.
In \textcolor{black}{Figure \ref{disp1}b}, the northern edge of the blue-shifted clump at ($l, \, b$)=(205\fdg35, -14\fdg30) coincides with the two B-type stars (HD~38563 and HD~290862) in NGC~2068. 
Besides, the B-type star (HD~290861) in NGC~2071 is also located at the northern edge of the blue-shifted cloud.
The three B-type stars, not only HD~38563 and HD~290862 in NGC~2068 but also HD~290861 in NGC~2071, tend to be located near the projected interface of the two clouds, and were possibly formed by a collisional trigger.}

\textcolor{black}{The projected displacement gives a collision timescale estimate, as shown by Fukui et al.  (2018b).}
On the basis of complementary distribution, we hypothesize that the two clouds collided with each other \textcolor{black}{$(0.8$ pc $\times \,2/\sqrt{3})/(1.5$ km s$^{-1}\,\times \, 2) = 0.3$ Myr} 
\textcolor{black}{ago at the latest, on a tentative assumption that the relative motion makes the same angle as that of the simulation (60$^\circ$) to the line of sight.}
This collision likely triggered formation of the three B-type stars and possibly the driving sources of the outflow as well as some of the YSOc near the interface regions of the two clouds.
The stars far from the interface were possibly formed prior to the collision.
The two B-type stars in NGC~2068 are located toward ($l, \, b$)=(205\fdg33, -14\fdg31), where a small CO clump overlaps after the displacement (Figure \ref{disp1}b) \textcolor{black}{in the scenario}.
The typical column density toward the collision spots is 10$^{22}$ cm$^{-2}$ in both components, similar to that found in other regions of single high-mass star formation by CCC (Fukui et al. 2018b).
\textcolor{black}{On the contrary, the B-type star in NGC~2071 is located at ($l, \, b$)=(205\fdg17, -14\fdg13), where the blue-shifted cloud and the intermediate cloud are dense while the red-shifted one is less dense as seen in Figure 4.}
A possible scenario is that the collision happens between two uneven clouds having column densities of 10$^{22}$ and 10$^{21}$ cm$^{-2}$ as found in NGC~6334 and NGC~6357 (Fukui et al. 2018a).
\textcolor{black}{In NGC~6334 and NGC~6357,} the boundary of the collisional area is not clearly specified toward the periphery of the clouds because part of the colliding clouds may be already dispersed by the collisional interaction.
Ionization is perhaps not so important, considering the present low ultraviolet radiation of B-type stars compared with that of O-type stars.
We also note that the directions of the outflow driving sources are possibly toward the regions of CCC.

\subsection{Large-scale star formation in Orion}
In the Orion region, the ionization-shock front driven by the H{\sc ii} regions was discussed above as a mechanism to form OB associations.
It is however difficult to understand the distribution of high-mass stars, O-type stars, and early B-type stars, lined up in Orion~A and Orion~B regions, with separations of $\sim$ 25 pc.
These large separations suggest that each instance of high-mass star formation is taking place independently and without mutual influence.
The present work showed that CCC provides a scenario to explain high-mass star formation in the NGC~2068 and NGC~2071 regions.
The process is similar to that presented for the M42/M43 region and the NGC~2024 \textcolor{black}{/NGC~2023} region.
In hydrodynamic simulations of isolated galaxies, the typical CCC timescale for a giant molecular cloud is several Myr (Fujimoto et al. 2014; Dobbs et al. 2015).
This is longer than that inferred for the three collisions in Orion within 1 Myr.
The three regions show displacements of 0.3--1.0 pc between the two components.
A global study connecting the individual regions is a future challenge to link the individual star formation with the galaxy scale gas motion.

\section{Conclusions}
We made new $^{13}$CO($J$=2--1) observations of the molecular cloud associated with two early B-type star-forming regions NGC~2068 and NGC~2071 in the L1630 dark cloud complex.
By analyzing the $^{13}$CO data we made detailed investigation of kinematical properties of the star forming gas. Conclusions are summarized as follows:

The northern cloud has an average velocity of \textcolor{black}{9.0} km s$^{-1}$ and the southern cloud an average velocity of \textcolor{black}{10.5} km s$^{-1}$. The two clouds are associated with early B-type star(s), protostellar outflow and several tens of YSOc, respectively.
We found that the two components show complementary distribution with each other; CO images of the blue-shifted and red-shifted clouds for a velocity interval of 1.5--2.0 km s$^{-1}$ produce a good complementary fit after a displacement of \textcolor{black}{0.8} pc.
\textcolor{black}{On the basis of a new numerical simulation and} this complementary distribution, we hypothesized that the two clouds collided with each other 0.3 Myr ago on a tentative assumption that the relative motion makes an angle of 60 $^\circ$ to the line of sight.
This collision likely triggered formation of the three B-type stars and the driving sources of the outflow as well as part of the YSOc.
The typical column density toward the collision spots is 10$^{22}$ cm$^{-2}$, and the variation of density in the initial clouds may control the stellar mass forming.
It is possible that a subset of YSOc was formed prior to the collision over a timescale of 1 Myr.
In the Orion region including M42, M43, NGC~2024, NGC~2068, and NGC~2071, \textcolor{black}{recent works including the present paper, most of the O-type/early B-type stars are explained by way of star formation triggered by CCC.}
This possibility offers a step forward to better understand high-mass star formation mechanisms and the role of CCC.

\section{Acknowledgments}
NANTEN2 project is an international collaboration of ten universities, Nagoya University, Osaka Prefecture University, University of Cologne, University of Bonn, Seoul National University, University of Chile, University of New South Wales, Macquarie University, University of Sydney, and Zurich Technical University. 
This work was financially supported by Grants-in-Aid for Scientific Research (KAKENHI) of the Japanese society for the Promotion of Science (JSPS; grant numbers 15H05694 and 15K17607).
In addition, the authors would like to thank Enago (www.enago.jp) for the English language review.

% ...

\newpage

\begin{figure}
 \begin{center}
  \includegraphics[width=\hsize]{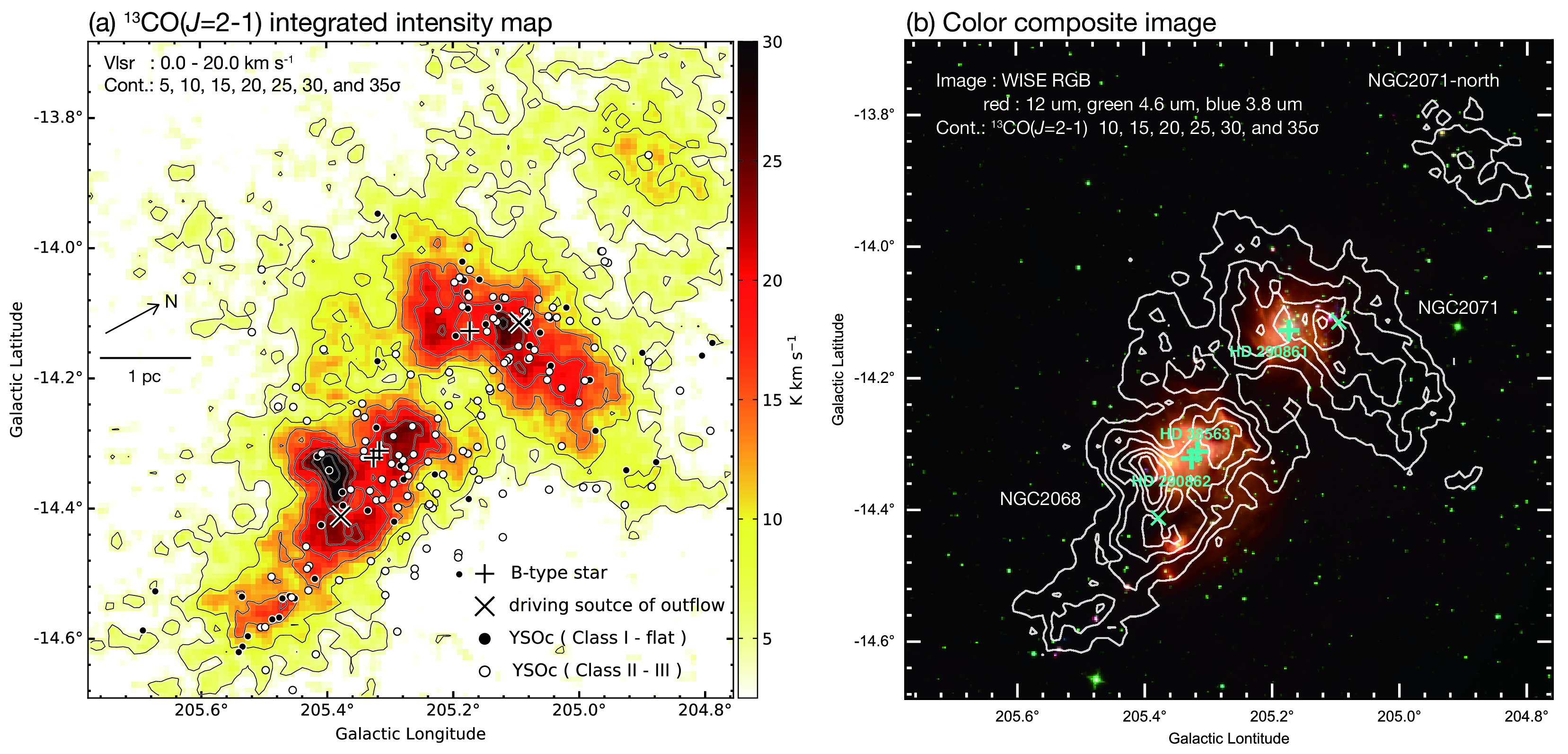}
 \end{center}
 \caption{
 \textcolor{black}{(a) The integrated intensity map of $^{13}$CO($J$=2--1) toward NGC~2068 and NGC~2071 integrated between 0.0 and 20.0 km s$^{-1}$.
 The black crosses and black saltire show B-type stars and driving source of outflow, respectively.
 The black and white circles represent Class I--Flat and Class II--III YSOc, respectively, listed in Spezzi et al. (2015).
 (b) Color composite image of  NGC~2068 and NGC~2071.
 Green, blue, and red color show WISE 12 $\mu$m, 4.6 $\mu$m, and 3.8 $\mu$m emissions, respectively.
 The contour shows the $^{13}$CO($J$=2--1) distribution same as in Figure 1a.
 }}\label{lb}
\end{figure}

\newpage

\begin{figure}
 \begin{center}
  \includegraphics[width=\hsize]{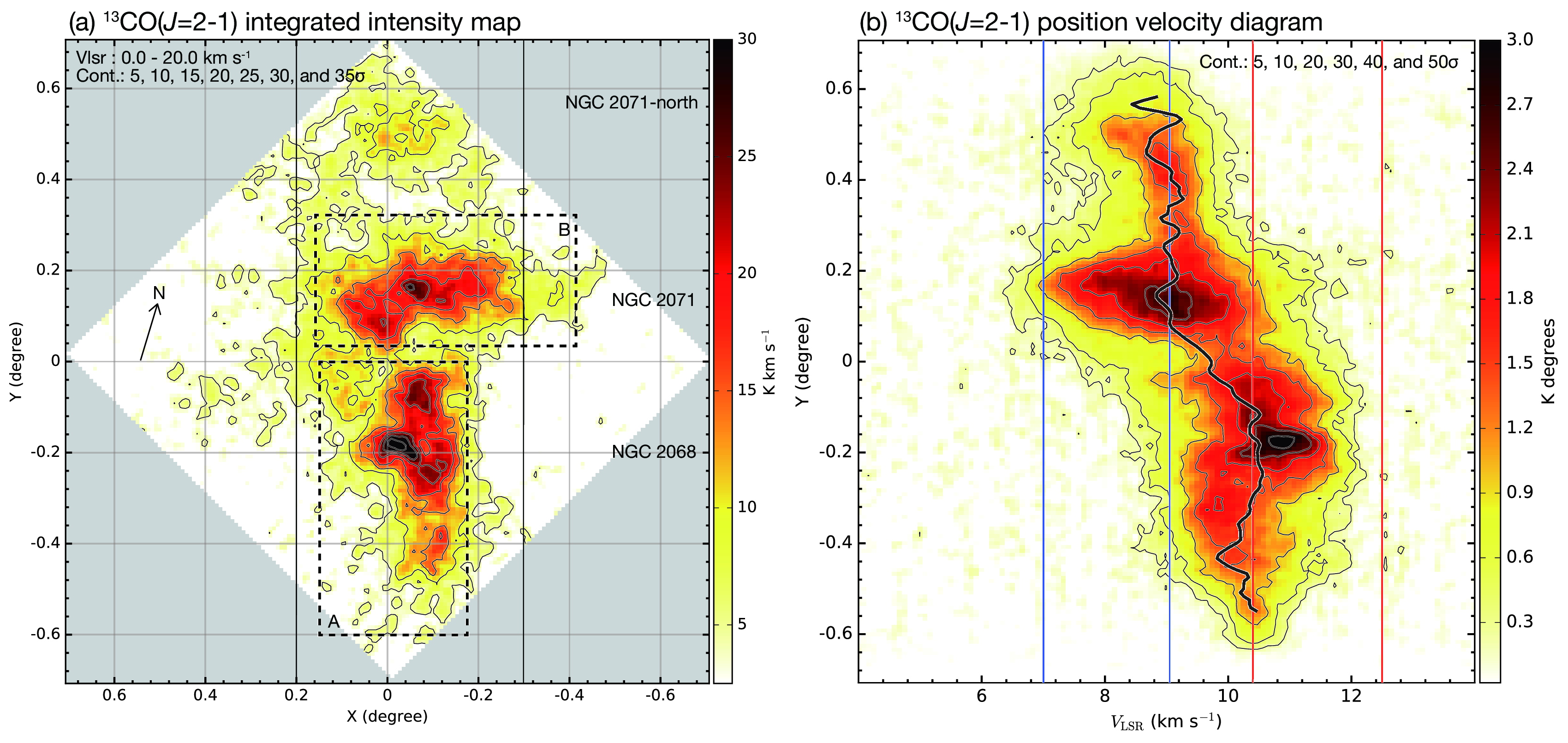}
 \end{center}
 \caption{
 \textcolor{black}{(a) The rotated integrated intensity map of $^{13}$CO($J$=2--1) toward NGC~2068 and NGC~2071 integrated between 0 and 20 km s$^{-1}$ (same as Figure1a) with a Y-axis in a position angle of 45$^{\circ}$ in Figure 1 and an X-axis orthogonal to the Y-axis, where (X,Y)=(0\fdg0, 0\fdg0) corresponds to ($l, \, b$)=(205\fdg26, -14\fdg19).}
 \textcolor{black}{The two black lines indicate the integration range of Figure 2b. Dashed boxes A and B indicate the areas to use for calculating the intensity-weighted mean velocity of north and south clouds.(b) The position-velocity diagram of $^{13}$CO($J$=2--1) emission toward NGC~2068 and NGC~2071.The black line shows the intensity-weighted mean velocity }
 \textcolor{black}{along the Y-axis at every 30$^{\prime \prime}$.}
 \textcolor{black}{The blue and red lines indicate the integration ranges of blue-shifted and red-shifted clouds presented in Figure 4.}
 }\label{pv}
\end{figure}

\newpage

\begin{figure}
 \begin{center}
  \includegraphics[width=\hsize]{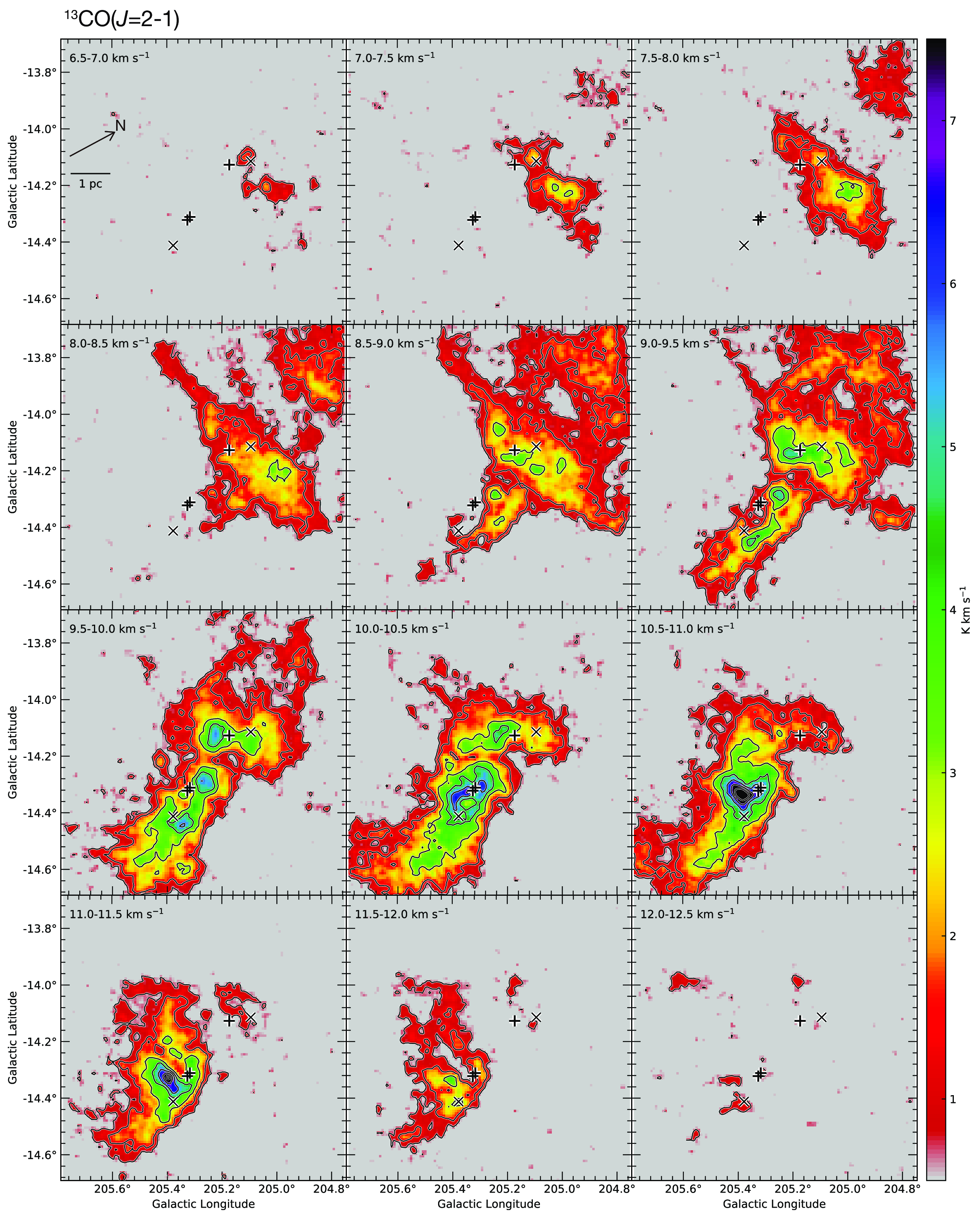}
 \end{center}
 \caption{
 \textcolor{black}{
 Channel maps of $^{13}$CO($J$=2--1) emission in 0.5 km s$^{-1}$ intervals over the velocity range from 6.5 to 12.5 km s$^{-1}$.
 The integration range is indicated in the top left corner of each panel.
 The crosses represent the positions of B-type stars.}
}\label{ch}
\end{figure}

\newpage

\begin{figure}
 \begin{center}
  \includegraphics[width=75mm]{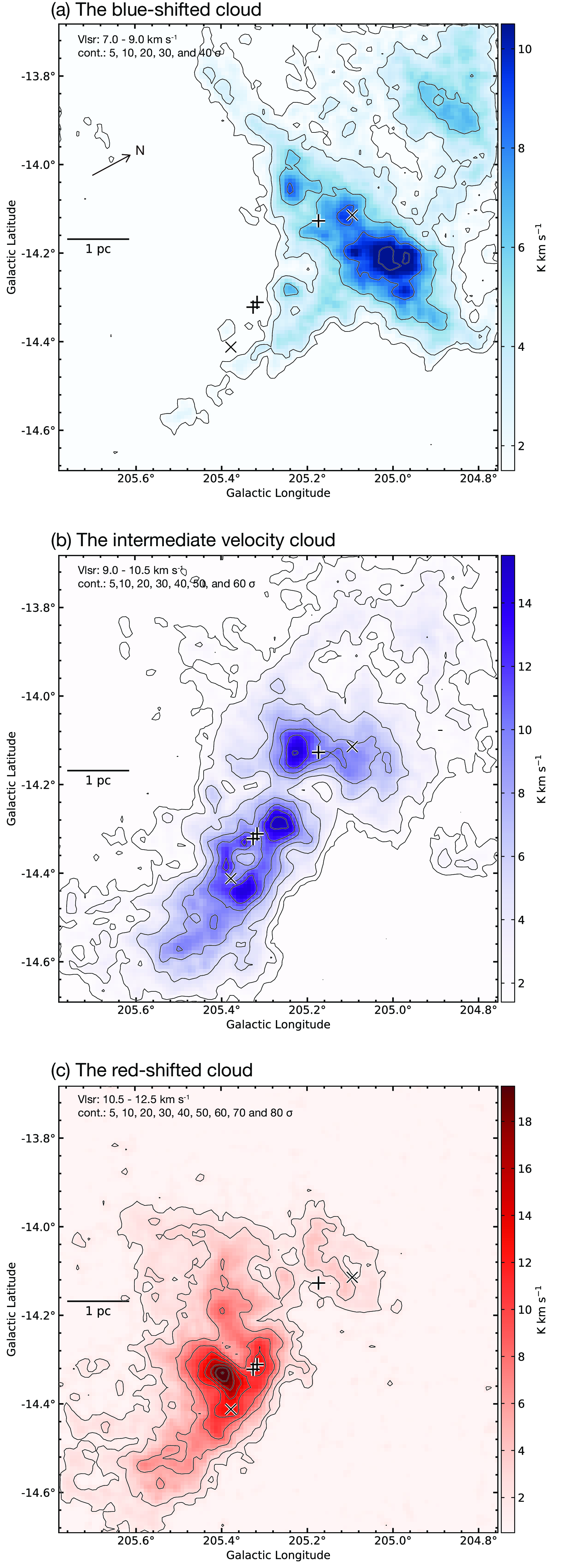}
 \end{center}
 \caption{
 \textcolor{black}{
 Spatial distribution of the three velocity components, 
 (a) blue-shifted cloud (7--9 km s$^{-1}$), 
 (b) intermediate velocity cloud (9.0--10.5 km s$^{-1}$), 
 (c) red-shifted cloud (10.5--12.5 km s$^{-1}$).
 The symbols are the same as in Figure 3a.
 }}\label{3v}
\end{figure}

\newpage

\begin{figure}
 \begin{center}
  \includegraphics[width=100mm]{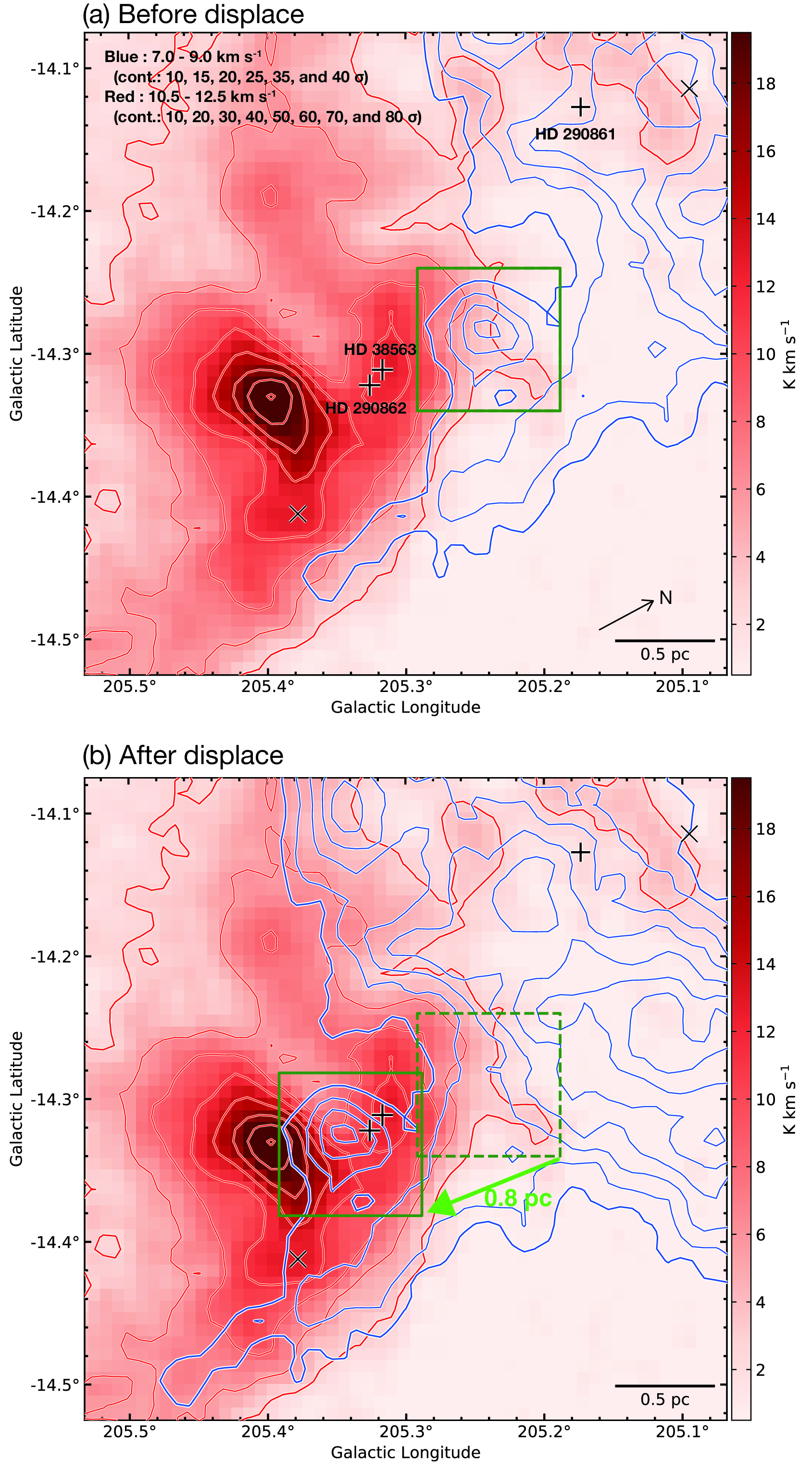}
 \end{center}
 \caption{
 \textcolor{black}{
 (a) Enlarged views of the gas distribution of the two velocity clouds.
 The image shows the$^{13}$CO($J$=2--1) intensity integrated over the velocity range of 10.5 km s$^{-1}$ to 12.5 km s$^{-1}$.
 The blue and red contours show the intensity integrated over the velocity range of 7.0--9.0 km s$^{-1}$ (blue) and 10.5--12.5 km s$^{-1}$ (red).
 The symbols are the same as in Figure 3a.
 (b) The blue-shifted cloud is displaced 0.8 pc as shown by the arrow. 
 }}\label{disp2}
\end{figure}

\newpage

\begin{figure}
 \begin{center}
  \includegraphics[width=140mm]{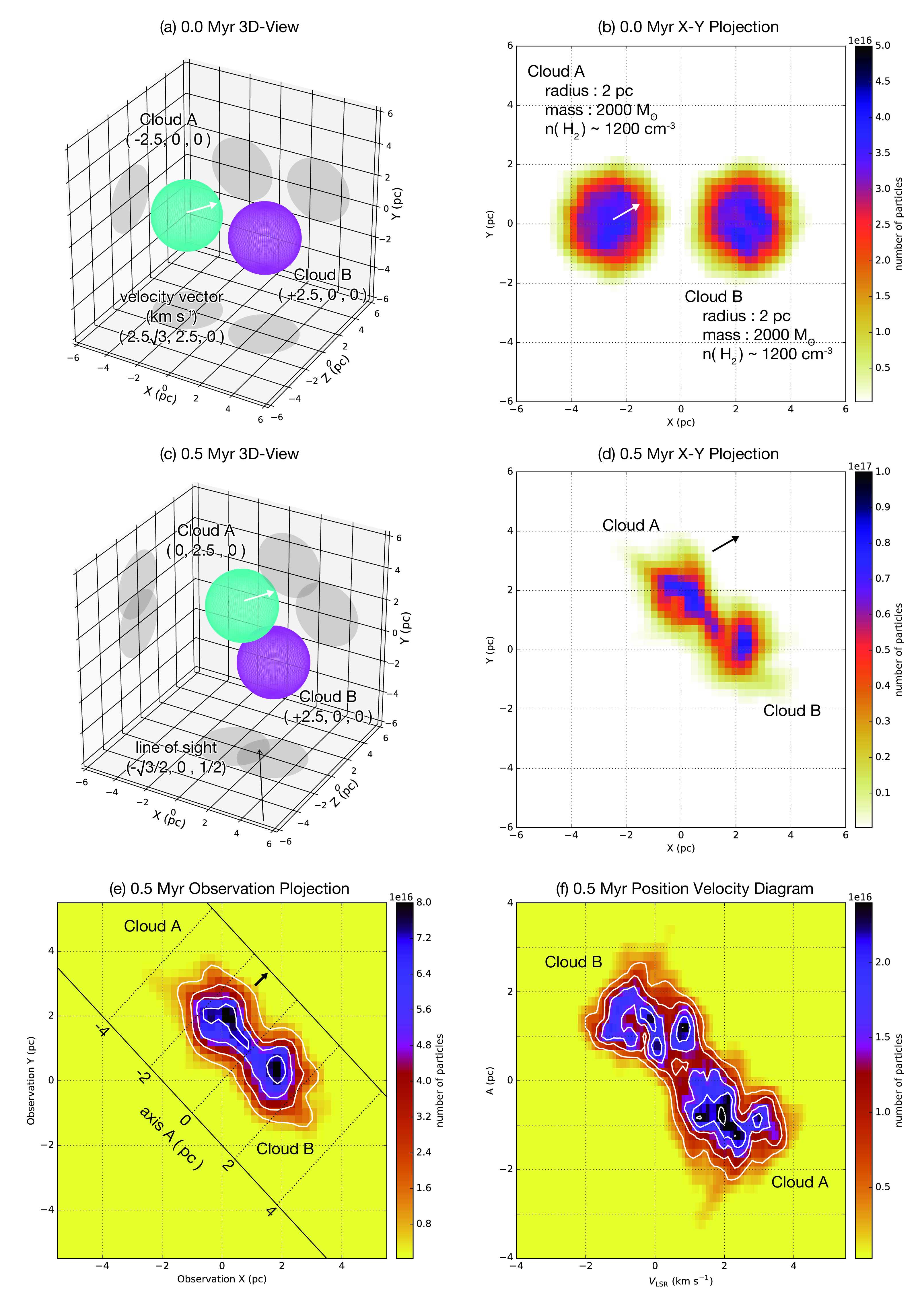}
 \end{center}
 \caption{
 \textcolor{black}{
 (a) The 3D schematic diagram of two colliding clouds in initial position.
 White arrow indicates the velocity vector of cloud A.
 (b) The initial position of clouds in X--Y projection.
 (c) The 3D schematic diagram of the two clouds at 0.5 Myr epoch.
 The black arrow shows the line of sight of Figures \ref{sim}e and \ref{sim}f.
 (d) The X--Y projection view of the clouds at 0.5 Myr.
 (e) The projected gas distribution at 0.5 Myr observed from 60$^\circ$.
 The two solid lines show the integration range in Figure \ref{sim}f.
 (f) The position-velocity diagram along the A--axis shown in Figure \ref{sim}e.
 }}\label{sim}
\end{figure}

\begin{figure}
 \begin{center}
  \includegraphics[width=100mm]{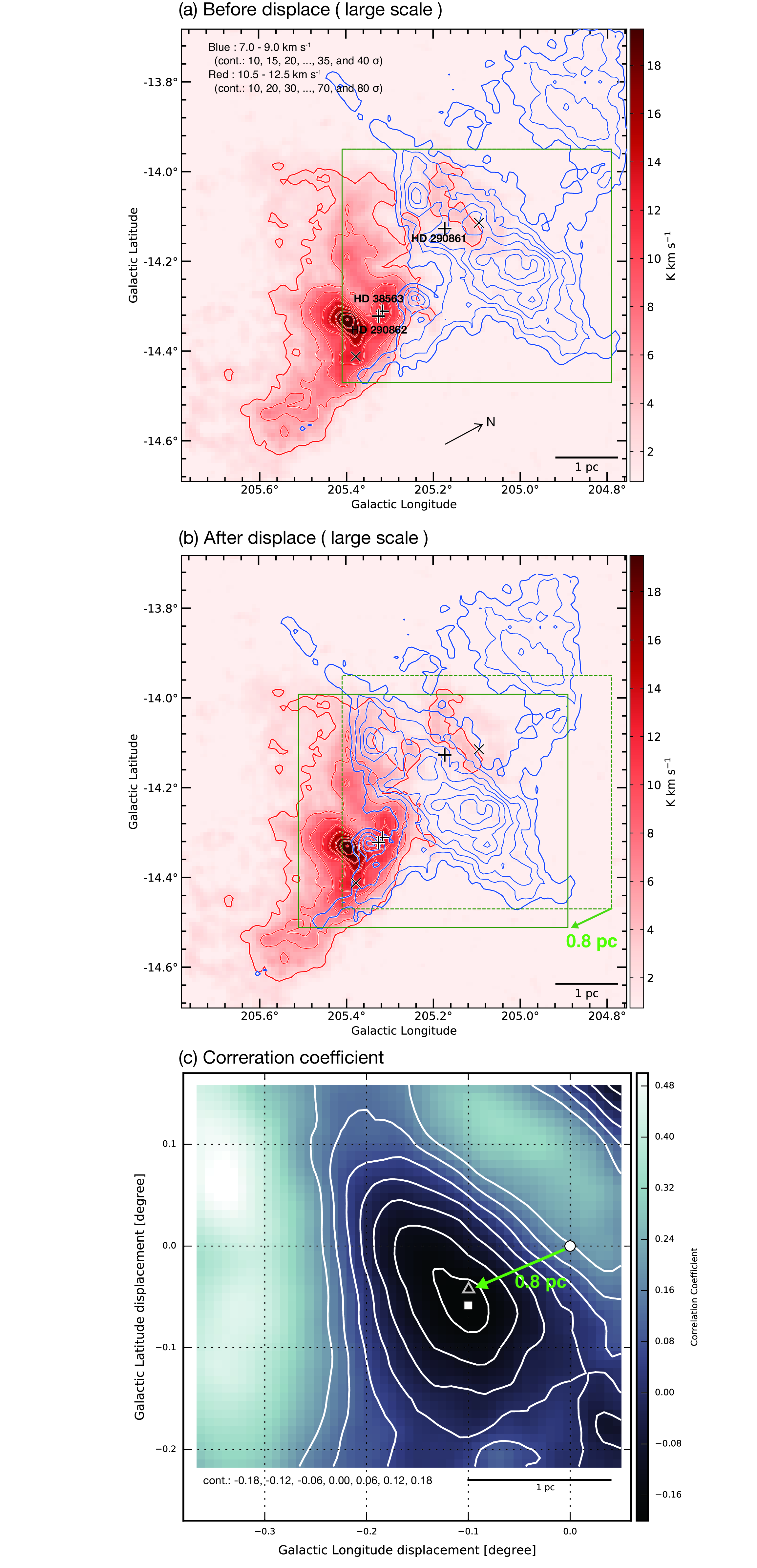}
 \end{center}
 \caption{
 \textcolor{black}{
 (a) The large scale gas distribution of the two velocity clouds as in Figure \ref{disp2}.
 (b) Same as (a), but the blue contours are displaced 0.8 pc as shown by the arrow.
 (c) The result from the correlation coefficient map.
 The white circles represent the initial position of the blue-shifted cloud, and the white square shows the weakest of correlation coefficient. The triangle represent the position of the displacement in Figures \ref{disp2}b and \ref{disp1}b. }
 }\label{disp1}
\end{figure}

\newpage

\newpage

 \begin{table}
   \tbl{The initial conditions of the numerical simulations.}{%
   \begin{tabular}{ccc}
   \hline\hline
   Simulation parameters & & \\ \hline
   Box size [pc] & 20 $\times$ 20 $\times$ 20 &   \\
   Resolution [pc] & 0.0039 & \\
   Ambient medium density [g cm$^{-3}$] & 1.69 $\times$ 10$^{-22}$ & \\
   Ambient medium temperature [K] & 800 & \\ \hline\hline
   Model parameters & Cloud A & Cloud B \\ \hline
   Initial Position ($x, y, z$) [pc] & (-2.5, 0, 0) & (+2.5, 0, 0)  \\
   Initial Velocity ($v_x, v_y, v_z$) [km s$^{-1}$]& ($2.5\sqrt{3}, \: 2.5, \: 0$) & (0, 0, 0)  \\
   Radius [pc] & 2 & 2 \\
   Mass [$M_\odot$] & 2000 & 2000 \\
   Initial Density Profile & Uniform & Uniform \\
   Free-fall time [Myr] & 1 & 1 \\
   Temperature [K] & 244.4 & 244.4 \\
   n$_{\rm H_2}$ [cm$^{-3}$ ] & 1217 & 1217 \\
   Density [g cm$^{-3}$] & 4.04 $\times$ 10$^{-21}$ & 4.04 $\times$ 10$^{-21}$ \\
   Velocity dispersion [km s$^{-1}$] & 1.53 & 1.53 \\ \hline
   \end{tabular}}\label{...}
   \begin{tabnote}
   \end{tabnote}
 \end{table}


\newpage

\end{document}